\documentclass[aps,preprint]{revtex4}%
\usepackage{amssymb}
\usepackage{amsmath}
\usepackage{epsfig}
\usepackage{graphicx}
\usepackage{amsfonts}%
\begin{document}
\title{High-energy Massive String Scatterings from Orientifold Planes}
\author{Jen-Chi Lee}
\email{jcclee@cc.nctu.edu.tw}
\affiliation{Department of Electrophysics, National Chiao-Tung University and Physics
Division, National Center for Theoretical Sciences, Hsinchu, Taiwan, R.O.C.}
\author{Yi Yang}
\email{yyang@phys.cts.nthu.edu.tw}
\affiliation{Department of Electrophysics, National Chiao-Tung University and Physics
Division, National Center for Theoretical Sciences, Hsinchu, Taiwan, R.O.C.}
\date{\today }

\begin{abstract}
We calculate bosonic massive closed string states at arbitrary mass levels
scattered from Orientifold planes in the high-energy, fixed angle limit. For
the case of O-particle scatterings, we obtain infinite linear relations among
high-energy scattering amplitudes of different string states. We also confirm
that there exist only closed string Regge poles in the form factor of the
O-particle amplitudes as expected. For the case of O-domain-wall scatterings,
we find that, like the well-known D-instanton scatterings, the amplitudes
behave like field theory scatterings, namely UV power-law without infinite
Regge poles. In addition, we discover that there exist only finite number of
$t$-channel closed string poles in the form factor of O-domain-wall
scatterings, and the masses of the poles are bounded by the masses of the
external legs. We thus confirm that all massive closed string states do couple
to the O-domain-wall.

\end{abstract}
\maketitle

\section{Introduction}

Being a consistent theory of quantum gravity, string theory is remarkable for
its soft ultraviolet structure. This is mainly due to two closely related
fundamental characteristics of high-energy string scattering amplitudes. The
first is the softer exponential fall-off behavior of the form factors of
high-energy string scatterings in contrast to the power-law (hard) behavior of
point particle field theory scatterings. The second is the existence of
infinite Regge poles in the form factor of string scattering amplitudes.
Recently high-energy, fixed angle string scattering amplitudes \cite{GM,
Gross, GrossManes} was reinvestigated for massive string states at arbitrary
mass levels \cite{ChanLee1,ChanLee2, CHL,CHLTY,PRL,paperB,susy,Closed, HL}. An
infinite number of linear relations among string scattering amplitudes were
obtained. The most important new ingredient of these calculations is the
zero-norm states (ZNS) \cite{ZNS1,ZNS3,ZNS2} in the old covariant first
quantized (OCFQ) string spectrum. The existence of these infinite linear
relations constitutes the \textit{third} fundamental characteristics of high
energy string scatterings. Other approaches related to this development can be
found in \cite{MooreWest}, 

These linear relations persist \cite{Dscatt} for string scattered from generic
D$p$-brane \cite{Klebanov} except D-instanton and D-domain-wall. For the
scattering of D-instanton, the form factor exhibits the well-known power-law
behavior without Regge pole structure, and thus resembles a field theory
amplitude. For the special case of D-domain-wall scattering \cite{Myers}, it
was discovered \cite{Wall} that its form factor behaves as\textit{ power-law}
with infinite \textit{open} Regge pole structure at high energies. This
discovery makes D-domain-wall scatterings an unique example of a hybrid of
string and field theory scatterings. Moreover, it was shown \cite{Wall} that
the linear relations break down for the D-domain-wall scattering due to this
unusual power-law behavior. This result seems to imply the coexistence of
linear relations and soft UV structure of string scatterings. Recent study of
high-energy scatterings of compatified closed string justified this conjecture
\cite{Compact}. In order to further uncover the mysterious relations among
these three fundamental characteristics of string scatterings, namely, the
soft UV structure, the existence of infinite Regge poles and the newly
discovered linear relations stated above, it will be important to study more
string scatterings, which exhibit the unusual behaviors in the high energy limit.

In this paper, we calculate massive closed string states at arbitrary mass
levels scattered from Orientifold planes in the high-energy, fixed angle
limit. The scatterings of massless states from Orientifold planes were
calculated previously by using the boundary states formalism
\cite{Craps,Schnitzer}, and more recently \cite{Garousi} on the worldsheet of
real projected plane $RP_{2}$. Many speculations were made about the
scatterings of \textit{massive} string states, in particular, for the case of
O-domain-wall scatterings. It is one of the purposes of this paper to clearify
these speculations and to discuss their relations with the three fundamental
characteristics of high-energy string scatterings stated above. For the
generic O$p$-planes with $p\geq0$, one expects to get the infinite linear
relations except O-domain-wall scatterings. For simplicity, we consider only
the case of O-particle scatterings. For the case of O-particle scatterings, we
obtain infinite linear relations among high-energy scattering amplitudes of
different string states. We also confirm that there exist only $t$-channel
closed string Regge poles in the form factor of the O-particle scatterings
amplitudes as expected. For the case of O-domain-wall scatterings, we find
that, like the well-known D-instanton scatterings, the amplitudes behave like
field theory scatterings, namely UV power-law without Regge pole. In addition,
we discover that there exist only finite number of $t$-channel closed string
poles in the form factor of O-domain-wall scatterings, and the masses of the
poles are bounded by the masses of the external legs. We thus confirm that all
massive closed string states do couple to the O-domain-wall as was conjectured
previously \cite{Myers, Garousi}. This is also consistent with the boundary
state descriptions of O-planes. For both cases of O-particle and O-domain-wall
scatterings, we confirm that there exist no $s$-channel open string Regge
poles in the form factor of the amplitudes as O-planes were known to be not
dynamical. However, the usual claim that there is a thinkness of
order$\sqrt{\alpha^{^{\prime}}}$ for the O-domain-wall is misleading as the UV
behavior of its scatterings is power-law instead of exponential fall-off. This
paper is organized as following. In section II, we write down a class of
high-energy vertex operators at general mass levels for the scatterings of
Orientifold planes. We then calculate the scattering from O-particle. In
section III, we calculate the scatterings from O-domain-wall and discuss the
pole structure in the form factor. A brief conclusion and discussion are given
in section IV.%

\setcounter{equation}{0}
\renewcommand{\theequation}{\arabic{section}.\arabic{equation}}%

\section{High-energy O-particle Scatterings}

We will use the real projected plane $RP_{2}$ as the worldsheet diagram for
the scatterings of Orientifold planes. The standard propagators of the left
and right moving fields are%
\begin{align}
\left\langle X^{\mu}\left(  z\right)  X^{\nu}\left(  w\right)  \right\rangle
&  =-\eta^{\mu\nu}\log\left(  z-w\right)  ,\label{D1}\\
\left\langle \tilde{X}^{\mu}\left(  \bar{z}\right)  \tilde{X}^{\nu}\left(
\bar{w}\right)  \right\rangle  &  =-\eta^{\mu\nu}\log\left(  \bar{z}-\bar
{w}\right)  . \label{D2}%
\end{align}
In addition, there are also nontrivial correlator between the right and left
moving fields as well%
\begin{equation}
\left\langle X^{\mu}\left(  z\right)  \tilde{X}^{\nu}\left(  \bar{w}\right)
\right\rangle =-D^{\mu\nu}\ln\left(  1+z\bar{w}\right)  . \label{DD}%
\end{equation}
As in the usual convention \cite{Klebanov}, the matrix $D$ reverses the sign
for fields satisfying Dirichlet boundary condition. The wave functions of a
tensor at general mass level can be written as%
\begin{equation}
T_{\mu_{1}\cdots\mu_{n}}=\dfrac{1}{2}\left[  \varepsilon_{\mu_{1}\cdots\mu
_{n}}e^{ik\cdot x}+\left(  D\cdot\varepsilon\right)  _{\mu_{1}}\cdots\left(
D\cdot\varepsilon\right)  _{\mu_{n}}e^{iD\cdot k\cdot x}\right]
\end{equation}
where%
\begin{equation}
\varepsilon_{\mu_{1}\cdots\mu_{n}}\equiv\varepsilon_{\mu_{1}}\cdots
\varepsilon_{\mu_{n}}.
\end{equation}
The vertex operators corresponding to the above wave functions are%
\begin{equation}
V\left(  \varepsilon,k,z,\bar{z}\right)  =\dfrac{1}{2}\left[  \varepsilon
_{\mu_{1}\cdots\mu_{n}}V^{\mu_{1}\cdots\mu_{n}}\left(  k,z,\bar{z}\right)
+\left(  D\cdot\varepsilon\right)  _{\mu_{1}}\cdots\left(  D\cdot
\varepsilon\right)  _{\mu_{n}}V^{\mu_{1}\cdots\mu_{n}}\left(  D\cdot
k,z,\bar{z}\right)  \right]  .
\end{equation}
For simplicity, we are going to calculate one tachyon and one massive closed
string state scattered from the O-particle in the high-energy limit. One
expects to get similar results for the generic O$p$-plane scatterings with
$p\geq0$ except O-domain-wall scatterings, which will be discussed in section
III. For this case $D_{\mu\nu}=-\delta_{\mu\nu}$, and the kinematic setup are%
\begin{align}
e^{P}  &  =\frac{1}{M}\left(  -E,-\mathrm{k}_{2},0\right)  =\frac{k_{2}}{M},\\
e^{L}  &  =\frac{1}{M}\left(  -\mathrm{k}_{2},-E,0\right)  ,\\
e^{T}  &  =\left(  0,0,1\right)  ,\\
k_{1}  &  =\left(  E,\mathrm{k}_{1}\cos\phi,-\mathrm{k}_{1}\sin\phi\right)
,\\
k_{2}  &  =\left(  -E,-\mathrm{k}_{2},0\right)
\end{align}
where $e^{P}$, $e^{L}$ and $e^{T}$ are polarization vectors of the tensor
state $k_{2}$ on the high-energy scattering plane. One can easily calculate
the following kinematic relations in the high-energy limit%
\begin{align}
e^{T}\cdot k_{2}  &  =e^{L}\cdot k_{2}=0,\\
e^{T}\cdot k_{1}  &  =-\mathrm{k}_{1}\sin\phi\sim-E\sin\phi,\\
e^{T}\cdot D\cdot k_{1}  &  =\mathrm{k}_{1}\sin\phi\sim E\sin\phi,\\
e^{T}\cdot D\cdot k_{2}  &  =0,\\
e^{L}\cdot k_{1}  &  =\frac{1}{M}\left[  \mathrm{k}_{2}E-\mathrm{k}_{1}%
E\cos\phi\right]  \sim\frac{E^{2}}{M}\left(  1-\cos\phi\right)  ,\\
e^{L}\cdot D\cdot k_{1}  &  =\frac{1}{M}\left[  \mathrm{k}_{2}E+\mathrm{k}%
_{1}E\cos\phi\right]  \sim\frac{E^{2}}{M}\left(  1+\cos\phi\right)  ,\\
e^{L}\cdot D\cdot k_{2}  &  =\frac{1}{M}\left[  -\mathrm{k}_{2}E-\mathrm{k}%
_{2}E\right]  \sim-\frac{2E^{2}}{M}.
\end{align}
We define%
\begin{align}
a_{0}  &  \equiv k_{1}\cdot D\cdot k_{1}=-E^{2}-\mathrm{k}_{1}^{2}\sim
-2E^{2},\\
a_{0}^{\prime}  &  \equiv k_{2}\cdot D\cdot k_{2}=-E^{2}-\mathrm{k}_{2}%
^{2}\sim-2E^{2},\\
b_{0}  &  \equiv k_{1}\cdot k_{2}=\left(  E^{2}-\mathrm{k}_{1}\mathrm{k}%
_{2}\cos\phi\right)  \sim E^{2}\left(  1-\cos\phi\right)  ,\\
c_{0}  &  \equiv k_{1}\cdot D\cdot k_{2}=\left(  E^{2}+\mathrm{k}%
_{1}\mathrm{k}_{2}\cos\phi\right)  \sim E^{2}\left(  1+\cos\phi\right)  ,
\end{align}
and the Mandelstam variables can be calculated to be%
\begin{align}
t  &  \equiv-\left(  k_{1}+k_{2}\right)  ^{2}=M_{1}^{2}+M_{2}^{2}-2k_{1}\cdot
k_{2}=M_{2}^{2}-2\left(  1+b_{0}\right)  ,\\
s  &  \equiv\dfrac{1}{2}k_{1}\cdot D\cdot k_{1}=\dfrac{1}{2}a_{0},\\
u  &  =-2k_{1}\cdot D\cdot k_{2}=-2c_{0}.
\end{align}
In the high-energy limit, we will consider an incoming tachyon state $k_{1}%
$and an outgoing tensor state $k_{2}$ of the following form%
\begin{equation}
\left(  \alpha_{-1}^{T}\right)  ^{n-m-2q}\left(  \alpha_{-2}^{L}\right)
^{q}\otimes\left(  \tilde{\alpha}_{-1}^{T}\right)  ^{n-m^{\prime}-2q^{\prime}%
}\left(  \tilde{\alpha}_{-2}^{L}\right)  ^{q^{\prime}}\left\vert
0\right\rangle .
\end{equation}
For simplicity, we have omited above a possible high-energy vertex
$(\alpha_{-1}^{L})^{r}\otimes(\tilde{\alpha}_{-1}^{L})^{r^{\prime}}$
\cite{Dscatt,Compact}. For this case, with momentum conservation on the
O-planes, we have%
\begin{equation}
a_{0}+b_{0}+c_{0}=M_{1}^{2}=-2. \label{conserve}%
\end{equation}
The high-energy scattering amplitude can then be written as
\begin{align*}
A^{RP_{2}}  &  =\int d^{2}z_{1}d^{2}z_{2}\dfrac{1}{2}\left[  V\left(
k_{1},z_{1}\right)  \tilde{V}\left(  k_{1},\bar{z}_{1}\right)  +V\left(
D\cdot k_{1},z_{1}\right)  \tilde{V}\left(  D\cdot k_{1},\bar{z}_{1}\right)
\right] \\
&  \cdot\dfrac{1}{2}\varepsilon_{T^{n-2q}L^{q},T^{n-2q^{\prime}}L^{q^{\prime}%
}}V^{T^{n-2q}L^{q}}\left(  k_{2},z_{2}\right)  \tilde{V}^{T^{n-2q^{\prime}%
}L^{q^{\prime}}}\left(  k_{2},\bar{z}_{2}\right) \\
&  +\left(  D\cdot\varepsilon_{T}\right)  ^{n-2q}\left(  D\cdot\varepsilon
_{L}\right)  ^{q}\left(  D\cdot\tilde{\varepsilon}_{T}\right)  ^{n-2q^{\prime
}}\left(  D\cdot\tilde{\varepsilon}_{L}\right)  ^{q^{\prime}}V^{T^{n-2q}L^{q}%
}\left(  D\cdot k_{2},z_{2}\right) \\
&  \cdot\tilde{V}^{T^{n-2q^{\prime}}L^{q^{\prime}}}\left(  D\cdot k_{2}%
,\bar{z}_{2}\right) \\
&  =A_{1}+A_{2}+A_{3}+A_{4}%
\end{align*}
where%
\begin{align}
A_{1}  &  =\dfrac{1}{4}\varepsilon_{T^{n-2q}L^{q},T^{n-2q^{\prime}%
}L^{q^{\prime}}}\int d^{2}z_{1}d^{2}z_{2}\nonumber\\
&  \cdot\left\langle V\left(  k_{1},z_{1}\right)  \tilde{V}\left(  k_{1}%
,\bar{z}_{1}\right)  V^{T^{n-2q}L^{q}}\left(  k_{2},z_{2}\right)  \tilde
{V}^{T^{n-2q^{\prime}}L^{q^{\prime}}}\left(  k_{2},\bar{z}_{2}\right)
\right\rangle ,\\
A_{2}  &  =\dfrac{1}{4}\varepsilon_{T^{n-2q}L^{q},T^{n-2q^{\prime}%
}L^{q^{\prime}}}\int d^{2}z_{1}d^{2}z_{2}\nonumber\\
&  \cdot\left\langle V\left(  D\cdot k_{1},z_{1}\right)  \tilde{V}\left(
D\cdot k_{1},\bar{z}_{1}\right)  V^{T^{n-2q}L^{q}}\left(  k_{2},z_{2}\right)
\tilde{V}^{T^{n-2q^{\prime}}L^{q^{\prime}}}\left(  k_{2},\bar{z}_{2}\right)
\right\rangle ,\\
A_{3}  &  =\dfrac{1}{4}\left(  D\cdot\varepsilon_{T}\right)  ^{n-2q}\left(
D\cdot\varepsilon_{L}\right)  ^{q}\left(  D\cdot\tilde{\varepsilon}%
_{T}\right)  ^{n-2q^{\prime}}\left(  D\cdot\tilde{\varepsilon}_{L}\right)
^{q^{\prime}}\nonumber\\
&  \cdot\int d^{2}z_{1}d^{2}z_{2}\left\langle V\left(  k_{1},z_{1}\right)
\tilde{V}\left(  k_{1},\bar{z}_{1}\right)  V^{T^{n-2q}L^{q}}\left(  D\cdot
k_{2},z_{2}\right)  \tilde{V}^{T^{n-2q^{\prime}}L^{q^{\prime}}}\left(  D\cdot
k_{2},\bar{z}_{2}\right)  \right\rangle ,\\
A_{4}  &  =\dfrac{1}{4}\left(  D\cdot\varepsilon_{T}\right)  ^{n-2q}\left(
D\cdot\varepsilon_{L}\right)  ^{q}\left(  D\cdot\tilde{\varepsilon}%
_{T}\right)  ^{n-2q^{\prime}}\left(  D\cdot\tilde{\varepsilon}_{L}\right)
^{q^{\prime}}\nonumber\\
&  \cdot\int d^{2}z_{1}d^{2}z_{2}\left\langle V\left(  D\cdot k_{1}%
,z_{1}\right)  \tilde{V}\left(  D\cdot k_{1},\bar{z}_{1}\right)
V^{T^{n-2q}L^{q}}\left(  D\cdot k_{2},z_{2}\right)  \tilde{V}^{T^{n-2q^{\prime
}}L^{q^{\prime}}}\left(  D\cdot k_{2},\bar{z}_{2}\right)  \right\rangle .
\end{align}
One can easily see that%
\begin{equation}
A_{1}=A_{4},A_{2}=A_{3}.
\end{equation}
We will choose to calculate $A_{1}$and $A_{2}$. For the case of $A_{1}$, we
have
\begin{align}
4A_{1}  &  =\varepsilon_{T^{n-2q}L^{q},T^{n-2q^{\prime}}L^{q^{\prime}}}\int
d^{2}z_{1}d^{2}z_{2}\cdot\nonumber\\
&  \left\langle e^{ik_{1}X}\left(  z_{1}\right)  e^{ik_{1}\tilde{X}}\left(
\bar{z}_{1}\right)  \left(  \partial X^{T}\right)  ^{n-2q}\left(
i\partial^{2}X^{L}\right)  ^{q}e^{ik_{2}X}\left(  z_{2}\right)  \left(
\bar{\partial}\tilde{X}^{T}\right)  ^{n-2q^{\prime}}\left(  i\bar{\partial
}^{2}\tilde{X}^{L}\right)  ^{q^{\prime}}e^{ik_{2}\tilde{X}}\left(  \bar{z}%
_{2}\right)  \right\rangle \nonumber\\
&  =\left(  -1\right)  ^{q+q^{\prime}}\int d^{2}z_{1}d^{2}z_{2}\left(
1+z_{1}\bar{z}_{1}\right)  ^{a_{0}}\left(  1+z_{2}\bar{z}_{2}\right)
^{a_{0}^{\prime}}\left\vert z_{1}-z_{2}\right\vert ^{2b_{0}}\left\vert
1+z_{1}\bar{z}_{2}\right\vert ^{2c_{0}}\nonumber\\
&  \cdot\left[  \frac{ie^{T}\cdot k_{1}}{z_{1}-z_{2}}-\frac{ie^{T}\cdot D\cdot
k_{1}}{1+\bar{z}_{1}z_{2}}\bar{z}_{1}-\frac{ie^{T}\cdot D\cdot k_{2}}%
{1+\bar{z}_{2}z_{2}}\bar{z}_{2}\right]  ^{n-2q}\nonumber\\
&  \cdot\left[  -\frac{ie^{T}\cdot D\cdot k_{1}}{1+z_{1}\bar{z}_{2}}%
z_{1}+\frac{ie^{T}\cdot k_{1}}{\bar{z}_{1}-\bar{z}_{2}}-\frac{ie^{T}\cdot
D\cdot k_{2}}{1+z_{2}\bar{z}_{2}}z_{2}\right]  ^{n-2q^{\prime}}\nonumber\\
&  \cdot\left[  \frac{e^{L}\cdot k_{1}}{\left(  z_{1}-z_{2}\right)  ^{2}%
}+\frac{e^{L}\cdot D\cdot k_{1}}{\left(  1+\bar{z}_{1}z_{2}\right)  ^{2}}%
\bar{z}_{1}^{2}+\frac{e^{L}\cdot D\cdot k_{2}}{\left(  1+\bar{z}_{2}%
z_{2}\right)  ^{2}}\bar{z}_{2}^{2}\right]  ^{q}\nonumber\\
&  \cdot\left[  \frac{e^{L}\cdot D\cdot k_{1}}{\left(  1+z_{1}\bar{z}%
_{2}\right)  ^{2}}z_{1}^{2}+\frac{e^{L}\cdot k_{1}}{\left(  \bar{z}_{1}%
-\bar{z}_{2}\right)  ^{2}}+\frac{e^{L}\cdot D\cdot k_{2}}{\left(  1+z_{2}%
\bar{z}_{2}\right)  ^{2}}z_{2}^{2}\right]  ^{q^{\prime}}.
\end{align}
To fix the modulus group on $RP_{2}$, choosing $z_{1}=r$ and $z_{2}=0$ and we
have%
\begin{align}
4A_{1}  &  =\left(  -1\right)  ^{n}\int_{0}^{1}dr^{2}\left(  1+r^{2}\right)
^{a_{0}}r^{2b_{0}}\nonumber\\
&  \cdot\left[  \frac{e^{T}\cdot k_{1}}{r}-\frac{e^{T}\cdot D\cdot k_{1}}%
{1}r\right]  ^{n-2q}\cdot\left[  -\frac{e^{T}\cdot D\cdot k_{1}}{1}%
r+\frac{e^{T}\cdot k_{1}}{r}\right]  ^{n-2q^{\prime}}\nonumber\\
&  \cdot\left[  \frac{e^{L}\cdot k_{1}}{r^{2}}+\frac{e^{L}\cdot D\cdot k_{1}%
}{1}r^{2}\right]  ^{q}\cdot\left[  \frac{e^{L}\cdot D\cdot k_{1}}{1}%
r^{2}+\frac{e^{L}\cdot k_{1}}{r^{2}}\right]  ^{q^{\prime}}\nonumber\\
&  =\left(  -1\right)  ^{n}\left(  E\sin\phi\right)  ^{2n}\left(  \frac
{2\cos^{2}\dfrac{\phi}{2}}{M\sin^{2}\phi}\right)  ^{q+q^{\prime}}\sum
_{i=0}^{q+q^{\prime}}\binom{q+q^{\prime}}{i}\left(  \dfrac{\sin^{2}\dfrac
{\phi}{2}}{\cos^{2}\dfrac{\phi}{2}}\right)  ^{i}\nonumber\\
&  \cdot\int_{0}^{1}dr^{2}\left(  1+r^{2}\right)  ^{a_{0}+2n-2\left(
q+q^{\prime}\right)  }\cdot\left(  r^{2}\right)  ^{b_{0}-n+2\left(
q+q^{\prime}\right)  -2i}.
\end{align}
Similarly, for the case of $A_{2}$, we have%
\begin{align}
4A_{2}  &  =\left(  -1\right)  ^{n}\int_{0}^{1}dr^{2}\left(  1+r^{2}\right)
^{a_{0}}r^{2c_{0}}\nonumber\\
&  \cdot\left[  \frac{e^{T}\cdot D\cdot k_{1}}{r}-\frac{e^{T}\cdot k_{1}}%
{1}r\right]  ^{n-2q}\cdot\left[  -\frac{e^{T}\cdot k_{1}}{1}r+\frac{e^{T}\cdot
D\cdot k_{1}}{r}\right]  ^{n-2q^{\prime}}\nonumber\\
&  \cdot\left[  \frac{e^{L}\cdot D\cdot k_{1}}{r^{2}}+\frac{e^{L}\cdot k_{1}%
}{1}r^{2}\right]  ^{q}\cdot\left[  \frac{e^{L}\cdot k_{1}}{1}r^{2}+\frac
{e^{L}\cdot D\cdot k_{1}}{r^{2}}\right]  ^{q^{\prime}}\nonumber\\
&  =\left(  -1\right)  ^{n}\left(  E\sin\phi\right)  ^{2n}\left(  \frac
{2\cos^{2}\dfrac{\phi}{2}}{M\sin^{2}\phi}\right)  ^{q+q^{\prime}}\sum
_{i=0}^{q+q^{\prime}}\binom{q+q^{\prime}}{i}\left(  \dfrac{\sin^{2}\dfrac
{\phi}{2}}{\cos^{2}\dfrac{\phi}{2}}\right)  ^{i}\nonumber\\
&  \cdot\int_{0}^{1}dr^{2}\text{ }\left(  1+r^{2}\right)  ^{a_{0}+2n-2\left(
q+q^{\prime}\right)  }\left(  r^{2}\right)  ^{c_{0}-n+2i}.
\end{align}
The scattering amplitude on $RP_{2}$ can therefore be calculated to be%
\begin{align}
A^{RP_{2}}  &  =A_{1}+A_{2}+A_{3}+A_{4}\nonumber\\
&  =\dfrac{1}{2}\left(  -1\right)  ^{n}\left(  E\sin\phi\right)  ^{2n}\left(
\frac{2\cos^{2}\dfrac{\phi}{2}}{M\sin^{2}\phi}\right)  ^{q+q^{\prime}}%
\sum_{i=0}^{q+q^{\prime}}\binom{q+q^{\prime}}{i}\left(  \dfrac{\sin^{2}%
\dfrac{\phi}{2}}{\cos^{2}\dfrac{\phi}{2}}\right)  ^{i}\nonumber\\
&  \cdot\int_{0}^{1}dr^{2}\left(  1+r^{2}\right)  ^{a_{0}+2n-2\left(
q+q^{\prime}\right)  }\cdot\left[  \left(  r^{2}\right)  ^{b_{0}-n+2\left(
q+q^{\prime}\right)  -2i}+\left(  r^{2}\right)  ^{c_{0}-n+2i}\right]  .
\label{RP2}%
\end{align}
The integral in Eq.(\ref{RP2}) can be calculated as following%
\begin{align}
&  \int_{0}^{1}dr^{2}\left(  1+r^{2}\right)  ^{a_{0}+2n-2\left(  q+q^{\prime
}\right)  }\cdot\left[  \left(  r^{2}\right)  ^{b_{0}-n+2\left(  q+q^{\prime
}\right)  -2i}+\left(  r^{2}\right)  ^{c_{0}-n+2i}\right] \nonumber\\
&  =[\dfrac{2^{1+a_{0}+2n-2\left(  q+q^{\prime}\right)  }}{1+b_{0}-n+2\left(
q+q^{\prime}\right)  -2i}]\nonumber\\
&  \cdot F\left(  2+a_{0}+b_{0}+n-2i,1,2+b_{0}-n+2\left(  q+q^{\prime}\right)
-2i,-1\right) \nonumber\\
&  +[\dfrac{2^{1+a_{0}+2n-2\left(  q+q^{\prime}\right)  }}{1+c_{0}%
-n+2i}]F\left(  2+a_{0}+c_{0}+n-2\left(  q+q^{\prime}\right)  +2i,1,2+c_{0}%
-n+2i,-1\right)
\end{align}
where we have used the following identities of the hypergeometric function
$F\left(  \alpha,\beta,\gamma,x\right)  $
\begin{align}
F(\alpha,\beta,\gamma;x)  &  =\frac{\Gamma(\gamma)}{\Gamma(\beta)\Gamma
(\gamma-\beta)}\int_{0}^{1}dy\text{ }y^{\beta-1}\left(  1-y\right)
^{\gamma-\beta-1}\left(  1-yx\right)  ^{-\alpha},\\
F\left(  \alpha,\beta,\gamma,x\right)   &  =2^{\gamma-\alpha-\beta}F\left(
\gamma-\alpha,\gamma-\beta,\gamma,x\right)  .
\end{align}
To further reduce the scattering amplitude into beta function, we use the
momentum conservation in Eq.(\ref{conserve}) and the identity%
\begin{align}
&  \left(  1+\alpha\right)  F\left(  -\alpha,1,2+\beta,-1\right)  +\left(
1+\beta\right)  F\left(  -\beta,1,2+\alpha,-1\right) \nonumber\\
&  =2^{1+\alpha+\beta}\dfrac{\Gamma\left(  \alpha+2\right)  \Gamma\left(
\beta+2\right)  }{\Gamma\left(  \alpha+\beta+2\right)  }%
\end{align}
to get
\begin{align}
\lbrack &  \dfrac{2^{1+a_{0}+2n-2\left(  q+q^{\prime}\right)  }}%
{1+b_{0}-n+2\left(  q+q^{\prime}\right)  -2i}]F\left(  -c_{0}+n-2i,1,2+b_{0}%
-n+2\left(  q+q^{\prime}\right)  -2i,-1\right) \nonumber\\
&  +[\dfrac{2^{1+a_{0}+2n-2\left(  q+q^{\prime}\right)  }}{1+c_{0}%
-n+2i}]F\left(  -b_{0}+n-2\left(  q+q^{\prime}\right)  +2i,1,2+c_{0}%
-n+2i,-1\right) \nonumber\\
&  =\dfrac{\Gamma\left(  1+c_{0}-n+2i\right)  \Gamma\left(  1+b_{0}-n+2\left(
q+q^{\prime}\right)  -2i\right)  }{\Gamma\left(  2+b_{0}+c_{0}-2n+2\left(
q+q^{\prime}\right)  \right)  }\nonumber\\
&  \sim B\left(  1+b_{0},1+c_{0}\right)  \dfrac{\left(  1+c_{0}\right)
^{-n+2i}\left(  1+b_{0}\right)  ^{-n+2\left(  q+q^{\prime}\right)  -2i}%
}{\left(  2+b_{0}+c_{0}\right)  ^{-2n+2\left(  q+q^{\prime}\right)  }%
}\nonumber\\
&  \sim B\left(  1+b_{0},1+c_{0}\right)  \left(  \cos^{2}\dfrac{\phi}%
{2}\right)  ^{-n+2i}\left(  \sin^{2}\dfrac{\phi}{2}\right)  ^{-n+2\left(
q+q^{\prime}\right)  -2i}.
\end{align}
We finally end up with%
\begin{align}
A^{RP_{2}}  &  =A_{1}+A_{2}+A_{3}+A_{4}\nonumber\\
&  =\dfrac{1}{2}\left(  -1\right)  ^{n}\left(  E\sin\phi\right)  ^{2n}\left(
\frac{2\cos^{2}\dfrac{\phi}{2}}{M\sin^{2}\phi}\right)  ^{q+q^{\prime}}%
\sum_{i=0}^{q+q^{\prime}}\binom{q+q^{\prime}}{i}\left(  \dfrac{\sin^{2}%
\dfrac{\phi}{2}}{\cos^{2}\dfrac{\phi}{2}}\right)  ^{i}\nonumber\\
&  \cdot B\left(  1+b_{0},1+c_{0}\right)  \left(  \cos^{2}\dfrac{\phi}%
{2}\right)  ^{-n+2i}\left(  \sin^{2}\dfrac{\phi}{2}\right)  ^{-n+2\left(
q+q^{\prime}\right)  -2i}\nonumber\\
&  =\dfrac{1}{2}\left(  -1\right)  ^{n}\left(  2E\right)  ^{2n}\left(
\frac{\sin^{2}\dfrac{\phi}{2}}{2M}\right)  ^{q+q^{\prime}}B\left(
1+b_{0},1+c_{0}\right)  \sum_{i=0}^{q+q^{\prime}}\binom{q+q^{\prime}}%
{i}\left(  \dfrac{\cos^{2}\dfrac{\phi}{2}}{\sin^{2}\dfrac{\phi}{2}}\right)
^{i}\nonumber\\
&  =\dfrac{1}{2}\left(  -1\right)  ^{n}\left(  2E\right)  ^{2n}\left(
\frac{1}{2M}\right)  ^{q+q^{\prime}}B\left(  1+b_{0},1+c_{0}\right)
\nonumber\\
&  \sim\dfrac{1}{2}\left(  -1\right)  ^{n}\left(  2E\right)  ^{2n}\left(
\frac{1}{2M}\right)  ^{q+q^{\prime}}B\left(  -\dfrac{t}{2},-\dfrac{u}%
{2}\right)  . \label{O-particle}%
\end{align}
From Eq.(\ref{O-particle}) we see that the UV behavior of O-particle
scatterings is exponential fall-off and one gets infinite linear relations
among string scattering amplitudes of different string states at each fixed
mass level. Note that both $t$ and $u$ correspond to the closed string channel
poles, while $s$ corresponds to the open string channel poles. It can be seen
from Eq.(\ref{O-particle}) that an infinite closed string Regge poles exist in
the form factor of O-particle scatterings. Furthermore, there are no
$s$-channel open string Regge poles as expected since O-planes are not
dynamical. This is in contrast to the D-particle scatterings \cite{Dscatt}
where both infinite $s$-channel open string Regge poles and $t$-channel closed
string Regge poles exist in the form factor. We will see that the fundamental
characteristics of O-domain-wall scatterings are very different from those of
O-particle scatterings as we will now discuss in the next section.

%

\setcounter{equation}{0}
\renewcommand{\theequation}{\arabic{section}.\arabic{equation}}%

\section{High-energy O-domain-wall Scatterings}

For this case the kinematic setup is%
\begin{align}
e^{P}  &  =\frac{1}{M}\left(  -E,\mathrm{k}_{2}\cos\theta,-\mathrm{k}_{2}%
\sin\theta\right)  =\frac{k_{2}}{M},\label{10}\\
e^{L}  &  =\frac{1}{M}\left(  -\mathrm{k}_{2},E\cos\theta,-E\sin\theta\right)
,\label{11}\\
e^{T}  &  =\left(  0,\sin\theta,\cos\theta\right)  ,\label{12}\\
k_{1}  &  =\left(  E,-\mathrm{k}_{1}\cos\phi,-\mathrm{k}_{1}\sin\phi\right)
,\label{13}\\
k_{2}  &  =\left(  -E,\mathrm{k}_{2}\cos\theta,-\mathrm{k}_{2}\sin
\theta\right)  . \label{14}%
\end{align}
In the high-energy limit, the angle of incidence $\phi$ is identical to the
angle of reflection $\theta$ and $Diag$ $D_{\mu\nu}=(-1,1,-1)$. The following
kinematic relations can be easily calculated%
\begin{align}
e^{T}\cdot k_{2}  &  =e^{L}\cdot k_{2}=0,\\
e^{T}\cdot k_{1}  &  =-2\mathrm{k}_{1}\sin\phi\cos\phi\sim-E\sin2\phi,\\
e^{T}\cdot D\cdot k_{1}  &  =0,\\
e^{T}\cdot D\cdot k_{2}  &  =2\mathrm{k}_{2}\sin\phi\cos\phi\sim E\sin2\phi,\\
e^{L}\cdot k_{1}  &  =\frac{1}{M}\left[  \mathrm{k}_{2}E-\mathrm{k}%
_{1}E\left(  \cos^{2}\phi-\sin^{2}\phi\right)  \right]  \sim\frac{2E^{2}}%
{M}\sin^{2}\phi,\\
e^{L}\cdot D\cdot k_{1}  &  =0,\\
e^{L}\cdot D\cdot k_{2}  &  =\frac{1}{M}\left[  -\mathrm{k}_{2}E+\mathrm{k}%
_{2}E\left(  \cos^{2}\phi-\sin^{2}\phi\right)  \right]  \sim-\frac{2E^{2}}%
{M}\sin^{2}\phi.
\end{align}
We define%
\begin{align}
a_{0}  &  \equiv k_{1}\cdot D\cdot k_{1}\sim-2E^{2}\sin^{2}\phi-2M_{1}^{2}%
\cos^{2}\phi+M_{1}^{2},\\
a_{0}^{\prime}  &  \equiv k_{2}\cdot D\cdot k_{2}=-E^{2}-\mathrm{k}_{2}%
^{2}\sim-2E^{2},\\
b_{0}  &  \equiv k_{1}\cdot k_{2}\sim2E^{2}\sin^{2}\phi+2M_{1}^{2}\cos^{2}%
\phi-\dfrac{1}{2}\left(  M_{1}^{2}+M^{2}\right)  ,\\
c_{0}  &  \equiv k_{1}\cdot D\cdot k_{2}=E^{2}-\mathrm{k}_{1}\mathrm{k}%
_{2}\sim\dfrac{1}{2}\left(  M_{1}^{2}+M^{2}\right)  , \label{c0}%
\end{align}
and the Mandelstam variables can be calculated to be%
\begin{align}
t  &  \equiv-\left(  k_{1}+k_{2}\right)  ^{2}=M_{1}^{2}+M_{2}^{2}-2k_{1}\cdot
k_{2}=M_{2}^{2}-2\left(  1+b_{0}\right)  ,\\
s  &  \equiv\dfrac{1}{2}k_{1}\cdot D\cdot k_{1}=\dfrac{1}{2}a_{0},\\
u  &  =-2k_{1}\cdot D\cdot k_{2}=-2c_{0}.
\end{align}
The first term of high-energy scatterings from O-domain-wall is%
\begin{align}
4A_{1}  &  =\left(  -1\right)  ^{n}\int_{0}^{1}dr^{2}\left(  1+r^{2}\right)
^{a_{0}}r^{2b_{0}}\nonumber\\
&  \cdot\left[  \frac{e^{T}\cdot k_{1}}{r}-\frac{e^{T}\cdot D\cdot k_{1}}%
{1}r\right]  ^{n-2q}\cdot\left[  -\frac{e^{T}\cdot D\cdot k_{1}}{1}%
r+\frac{e^{T}\cdot k_{1}}{r}\right]  ^{n-2q^{\prime}}\nonumber\\
&  \cdot\left[  \frac{e^{L}\cdot k_{1}}{r^{2}}+\frac{e^{L}\cdot D\cdot k_{1}%
}{1}r^{2}\right]  ^{q}\cdot\left[  \frac{e^{L}\cdot D\cdot k_{1}}{1}%
r^{2}+\frac{e^{L}\cdot k_{1}}{r^{2}}\right]  ^{q^{\prime}}\nonumber\\
&  \sim\left(  -1\right)  ^{n}\left(  E\sin2\phi\right)  ^{2n}\left(  \frac
{1}{2M\cos^{2}\phi}\right)  ^{q+q^{\prime}}\int_{0}^{1}dr^{2}\left(
1+r^{2}\right)  ^{a_{0}}\left(  r^{2}\right)  ^{b_{0}-n}.
\end{align}
The second term can be similarly calculated to be%
\begin{align}
4A_{2}  &  =\left(  -1\right)  ^{n}\int_{0}^{1}dr^{2}\left(  1+r^{2}\right)
^{a_{0}}r^{2c_{0}}\nonumber\\
&  \cdot\left[  \frac{e^{T}\cdot D\cdot k_{1}}{r}-\frac{e^{T}\cdot k_{1}}%
{1}r\right]  ^{n-2q}\cdot\left[  -\frac{e^{T}\cdot k_{1}}{1}r+\frac{e^{T}\cdot
D\cdot k_{1}}{r}\right]  ^{n-2q^{\prime}}\nonumber\\
&  \cdot\left[  \frac{e^{L}\cdot D\cdot k_{1}}{r^{2}}+\frac{e^{L}\cdot k_{1}%
}{1}r^{2}\right]  ^{q}\cdot\left[  \frac{e^{L}\cdot k_{1}}{1}r^{2}+\frac
{e^{L}\cdot D\cdot k_{1}}{r^{2}}\right]  ^{q^{\prime}}\nonumber\\
&  \sim\left(  -1\right)  ^{n}\left(  E\sin2\phi\right)  ^{2n-2\left(
q+q^{\prime}\right)  }\left(  \frac{2E^{2}}{M}\sin^{2}\phi\right)
^{q+q^{\prime}}\int_{0}^{1}dr^{2}\text{ }\left(  1+r^{2}\right)  ^{a_{0}%
}\left(  r^{2}\right)  ^{c_{0}+n}.
\end{align}
The scattering amplitudes of O-domain-wall on $RP_{2}$ can therefore be
calculated to be%
\begin{align}
A^{RP_{2}}  &  =A_{1}+A_{2}+A_{3}+A_{4}\nonumber\\
&  =\dfrac{1}{2}\left(  -1\right)  ^{n}\left(  E\sin2\phi\right)  ^{2n}\left(
\frac{1}{2M\cos^{2}\phi}\right)  ^{q+q^{\prime}}\nonumber\\
&  \cdot\int_{0}^{1}dr^{2}\left(  1+r^{2}\right)  ^{a_{0}}\left[  \left(
r^{2}\right)  ^{b_{0}-n}+\left(  r^{2}\right)  ^{c_{0}+n}\right]  .
\end{align}
By using the similar techanique for the case of O-particle scatterings, the
integral above can be calculated to be%
\begin{align}
&  \int dr^{2}\left(  1+r^{2}\right)  ^{a_{0}}\left[  \left(  r^{2}\right)
^{b_{0}-n}+\left(  r^{2}\right)  ^{c_{0}+n}\right] \nonumber\\
&  =\dfrac{F\left(  -a_{0},1+b_{0}-n,2+b_{0}-n,-1\right)  }{1+b_{0}-n}%
+\dfrac{F\left(  -a_{0},1+c_{0}+n,2+c_{0}+n,-1\right)  }{1+c_{0}+n}\nonumber\\
&  =\dfrac{2^{2+a_{0}+b_{0}+c_{0}}}{\left(  1+b_{0}-n\right)  \left(
1+c_{0}+n\right)  }\dfrac{\Gamma\left(  2+c_{0}+n\right)  \Gamma\left(
2+b_{0}-n\right)  }{\Gamma\left(  2+b_{0}+c_{0}\right)  }\nonumber\\
&  =\dfrac{\Gamma\left(  1+c_{0}+n\right)  \Gamma\left(  1+b_{0}-n\right)
}{\Gamma\left(  2+b_{0}+c_{0}\right)  }.
\end{align}
One thus ends up with%
\begin{align}
A^{RP_{2}}  &  =A_{1}+A_{2}+A_{3}+A_{4}\nonumber\\
&  =\dfrac{1}{2}\left(  -1\right)  ^{n}\left(  E\sin2\phi\right)  ^{2n}\left(
\frac{1}{2M\cos^{2}\phi}\right)  ^{q+q^{\prime}}\dfrac{\Gamma\left(
c_{0}+n+1\right)  \Gamma\left(  b_{0}-n+1\right)  }{\Gamma\left(  b_{0}%
+c_{0}+2\right)  }. \label{pole}%
\end{align}
Some crucial points of this result are in order. First, since $c_{0}$ is a
constant in the high-energy limit, the UV behavior of the O-domain-wall
scatterings is power-law instead of the usual exponential fall-off in other
O-plane scatterings. Second, there exist only \textit{finite} number of closed
string poles in the form factor. Note that although we only look at the high
energy kinimatic regime of the scattering amplitudes, it is easy to see that
there exists no infinite closed string Regge poles in the scattering
amplitudes for the whole kinematic regime. This is because there is only one
kinematic variable for the O-domain-wall scatterings. In fact, the structure
of poles in Eq$.$(\ref{pole}) can be calculated to be
\begin{align}
&  \dfrac{\Gamma\left(  1+c_{0}+n\right)  \Gamma\left(  1+b_{0}-n\right)
}{\Gamma\left(  2+b_{0}+c_{0}\right)  }\nonumber\\
&  =\dfrac{\Gamma\left(  1+M^{2}\right)  \Gamma\left(  1+b_{0}-n\right)
}{\Gamma\left(  b_{0}+n\right)  }\nonumber\\
&  =\Gamma\left(  1+M^{2}\right)  \dfrac{\left(  b_{0}-n\right)  !}{\left(
b_{0}+n-1\right)  !}\nonumber\\
&  =\Gamma\left(  1+M^{2}\right)
{\displaystyle\prod\limits_{k=1-n}^{n-1}}
\dfrac{1}{b_{0}-k}%
\end{align}
where we have used $c_{0}\equiv\dfrac{1}{2}\left(  M_{1}^{2}+M^{2}\right)  $
in the high-energy limit. It is easy to see that the larger the mass $M$ of
the external leg is, the more numerous the closed string poles are. We thus
confirm that all massive string states do couple to the O-domain-wall as was
conjectured previously \cite{Myers, Garousi}. This is also consistent with the
boundary state descriptions of O-planes. However, the claim that there is a
thinkness of order$\sqrt{\alpha^{^{\prime}}}$ for the O-domain-wall is
misleading as the UV behavior of its scatterings is power-law instead of
exponential fall-off. This concludes that, in contrast to the usual behavior
of high-energy, fixed angle string scattering amplitudes, namely soft UV,
linear relations and the existence of infinite Regge poles, O-domain-wall
scatterings, like the well-known D-instanton scatterings, behave like field
theory scatterings.

\section{Conclusions and discussions}%

\begin{figure}
[ptb]
\begin{center}
\includegraphics[
height=2.6247in,
width=5.047in
]%
{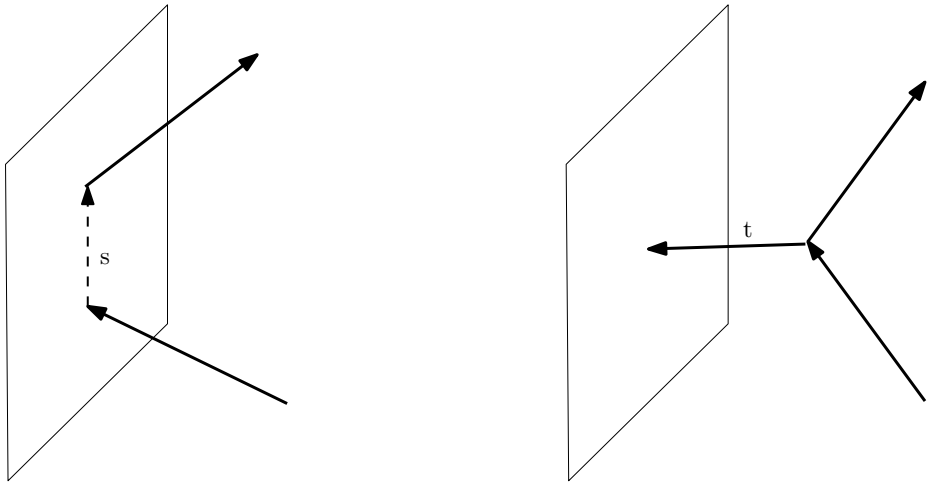}%
\caption{There are two possible channals for closed strings scattered from
D-branes/O-planes. The diagram on the left hand side corresponds to the
s-channel scatterings, and the diagram on the right hand side is the t-channel
scatterings.}%
\label{t-s}%
\end{center}
\end{figure}

In this paper, we calculate bosonic massive closed string states at arbitrary
mass levels scattered from Orientifold planes in the high-energy limit. We
have concentrated on the discussions of three fundamental characteristics of
high-energy, fixed angle string scattering amplitudes, namely soft UV,
infinite Regge poles and infinite linear relations discovered recently. For
the case of O-particle scatterings, we obtain infinite linear relations among
high-energy scattering amplitudes of different string states at each fixed
mass level. Moreover, the amplitude was found to be UV soft, namely,
exponential fall-off behavior. We also confirm that there exist only infinite
$t$-channel closed string Regge poles in the form factor of the O-particle
scatterings amplitudes as expected. For the case of O-domain-wall scatterings,
we find that, like the well-known D-instanton scatterings, the amplitudes
behave like field theory scatterings, namely UV power-law without infinite
Regge poles. In addition, we discover that there exist only finite number of
$t$-channel closed string poles in the form factor, and the masses of the
poles are bounded by the masses of the external legs. We thus confirm that all
massive closed string states do couple to the O-domain-wall as was conjectured
previously \cite{Myers, Garousi}. This is also consistent with the boundary
state descriptions of O-planes. For both cases of O-particle and O-domain-wall
scatterings, we confirm that there exist no open string Regge poles in the
form factor of the amplitudes as O-planes were known to be not dynamical.

We summarize the Regge pole structures of closed strings states scattered from
various D-branes and O-planes in the table. The $s$-channel and $t$-channel
scatterings for both D-branes and O-planes are shown in the Fig. 1. For
O-plane scatterings, the $s$-channel open string Regge poles are not allowed
since O-planes are not dynamical. For both cases of Domain-wall scatterings,
the $t$-channel closed string Regge poles are not allowed since there is only
one kinematic variable instead of two as in the usual cases.

\begin{center}
\ \
\begin{tabular}
[c]{|c|c|c|c|}\hline
& $p=-1$ & $1\leq p\leq23$ & $p=24$\\\hline
D$p$-branes & X & C+O & O\\\hline
O$p$-planes & X & C & X\\\hline
\end{tabular}

\end{center}

In this table, "C" and "O" represent infinite Closed string Regge poles and
Open string Regge poles respectively. "X" means there are no infinite Regge poles.

\section{Acknowledgments}

We would like to thank the hospitality of University of Tokyo at Komaba, where
most of this work was done. We are indebted to Prof. Tamiaki Yoneya for many
of his enlightening discussions. This work is supported in part by the
National Science Council, 50 billions project of Ministry of Educaton and
National Center for Theoretical Science, Taiwan, R.O.C.

\end{document}